\documentclass{article}
\usepackage{spconf,amsmath,graphicx}
\usepackage{diagbox}
\usepackage{graphicx} 
\usepackage{float} 
\usepackage{subfigure} 
\usepackage{caption}
\usepackage{color}

\title{DasFormer: deep alternating spectrogram transformer for multi/single-channel speech separation}
\name{Shuo Wang$^{1,2}$\sthanks{This work was done at Microsoft Research Asia.}, Xiangyu Kong$^2$\sthanks{Corresponding author (xiakon@microsoft.com).}, Xiulian Peng$^2$, Mahmood Movassagh$^3$, Vinod Prakash$^3$, Yan Lu$^2$}
\address{$^1$Institute of Acoustics, Chinese Academy of Sciences, Beijing, China \\
	$^2$Microsoft Research Asia, Beijing, China \\
	$^3$Microsoft Corporation, Redmond WA 98052, USA}

\begin{document}
	%
	\maketitle
	\begin{abstract}
		
		
		
		For the task of speech separation, previous study usually treats multi-channel and single-channel scenarios as two research tracks with specialized solutions developed respectively. Instead, we propose a simple and unified architecture - DasFormer (\textbf{D}eep \textbf{a}lternating \textbf{s}pectrogram trans\textbf{Former}) to handle both of them in the challenging reverberant environments. Unlike frame-wise sequence modeling, each TF-bin in the spectrogram is assigned with an embedding encoding spectral and spatial information. 
		With such input, DasFormer is then formed by multiple repetition of simple blocks each of which integrates \textit{1) two multi-head self-attention (MHSA) modules alternately processing within each frequency bin \& temporal frame of the spectrogram 2) MBConv before each MHSA for modeling local features on the spectrogram}. Experiments show that DasFormer has a powerful ability to model the time-frequency representation, whose performance far exceeds the current SOTA models in multi-channel speech separation, and also achieves single-channel SOTA in the more challenging yet realistic reverberation scenario. 
		
		
		
		
		
	\end{abstract}
	
	\section{Introduction}
	
	Deep neural networks (DNNs) based speech separation systems have received widespread attention since Deep Clustering (DC) \cite{hershey2016deep} and Permutation Invariant Training (PIT) \cite{yu2017permutation} were proposed. When multi-microphone data are available, many approaches merging spatial cues with deep models successively achieve SOTA results on multi-channel separation tasks\cite{wang2018multi, quan22b_interspeech, chen22e_interspeech}. However, the majority of the methods are specially designed for each scenario where cross-generalization usually becomes suboptimal. By contrast, in this paper we explore a network backbone capable of handling both single/multi-channel speech separation well.
	
	
	
	Early studies on speech separation mainly rely on the sparsity of speech in the time-frequency (TF) domain. This process can be performed by assigning speaker label to each TF-bin \cite{hershey2016deep}, either estimating a ratio mask and product with the original input, or directly estimating the complex coefficients \cite{wang2020complex}.
	Later, TasNet \cite{luo2018tasnet} and its variants \cite{luo2019conv,luo2020dual,chen2020dual} have grown to be dominant for speech separation under anechoic conditions. This often attributes to a learnable analytic basis instead of the fixed Fourier basis \cite{luo2018tasnet} and a more powerful network structure like Transformer \cite{chen2020dual, subakan2021attention, yang2022tfpsnet}. These separation methods have achieved satisfactory performance in anechoic environments, while degradation is evident when room reverberation is not neglected \cite{maciejewski2020whamr, tzinis2022compute}. Besides, the dual-path sequence modeling way like \cite{chen2020dual} has also gained attention in related fields such as speech enhancement \cite{dang2022dpt}.
	
	
	When multi-microphone data are available, the task falls into multi-channel speech separation. A remarkably successful solution \cite{ochiai2020beam} is to combine an optimal beamformer with a neural network (NN) like TasNet. 
	Subsequently, this cascade design was extended to the iterative refinement framework with improved performance \cite{chen22e_interspeech}. 
	Another common approach is end-to-end network design, which attempts to incorporate multi-channel cues such as Inter-channel Phase Difference (IPD) into the input of the network \cite{wang2018multi}. Recently, narrow-band conformer (NBC) \cite{quan22b_interspeech} handles speech separation with impressive gains in a narrow-band mode where all sub-bands share the same parameters.
	Note that valuable spatial information even exists for single-channel separation due to the presence of room reflections and reverberations \cite{patterson22_interspeech}.
	
	
	
	As far as the authors view, few efforts have been made on a common backbone network handling both tasks well. For instance, Sepformer\cite{subakan2021attention}, which performs better than TasNet in single-channel, does not work as well as Beam-TasNet when multiple microphones are available, and degrades severely in reverberant environments. Similarly, multi-channel methods like NBC \cite{quan22b_interspeech} surely handle reverberation cases, but they perform even worse than majority of single-channel systems when less channels are available. Besides, some multi-channel architectures are customized, which becomes incompatible with single-channel case. 
	
	In this paper, we argue that with proper modeling and deeper backbone networks, embedding with minimum units in TF-bins can encode enough information to separate speakers. Such embedding includes the dependencies of one TF unit with its surroundings and even distant area, and the spatial information introduced in the reverberant environment or/and multi-channel sampling. Specifically, 
	spectrogram-like features are then alternately fed into a block including two convolution modules, a spectral attention module and a temporal attention module. In this way, we can facilitate the aggregation of stable local features, spectral features and spatial features including direct path and reverberation respectively. Furthermore, by repeating such processing block, a deeper network - DasFormer allows each TF embedding to gather global clues from distant frames/sub-bands. 
	
	Experiments show that DasFormer achieves scale-invariant signal-to-distortion ratio improvement (SI-SDRi) far exceeding SOTA both on the spatialized WSJ0-2mix dataset (multi-channel) and a challenging single-channel dataset WHAMR! with room reverberation.

	\begin{figure}[t]
		\centering
		\includegraphics[width=0.4\textwidth]{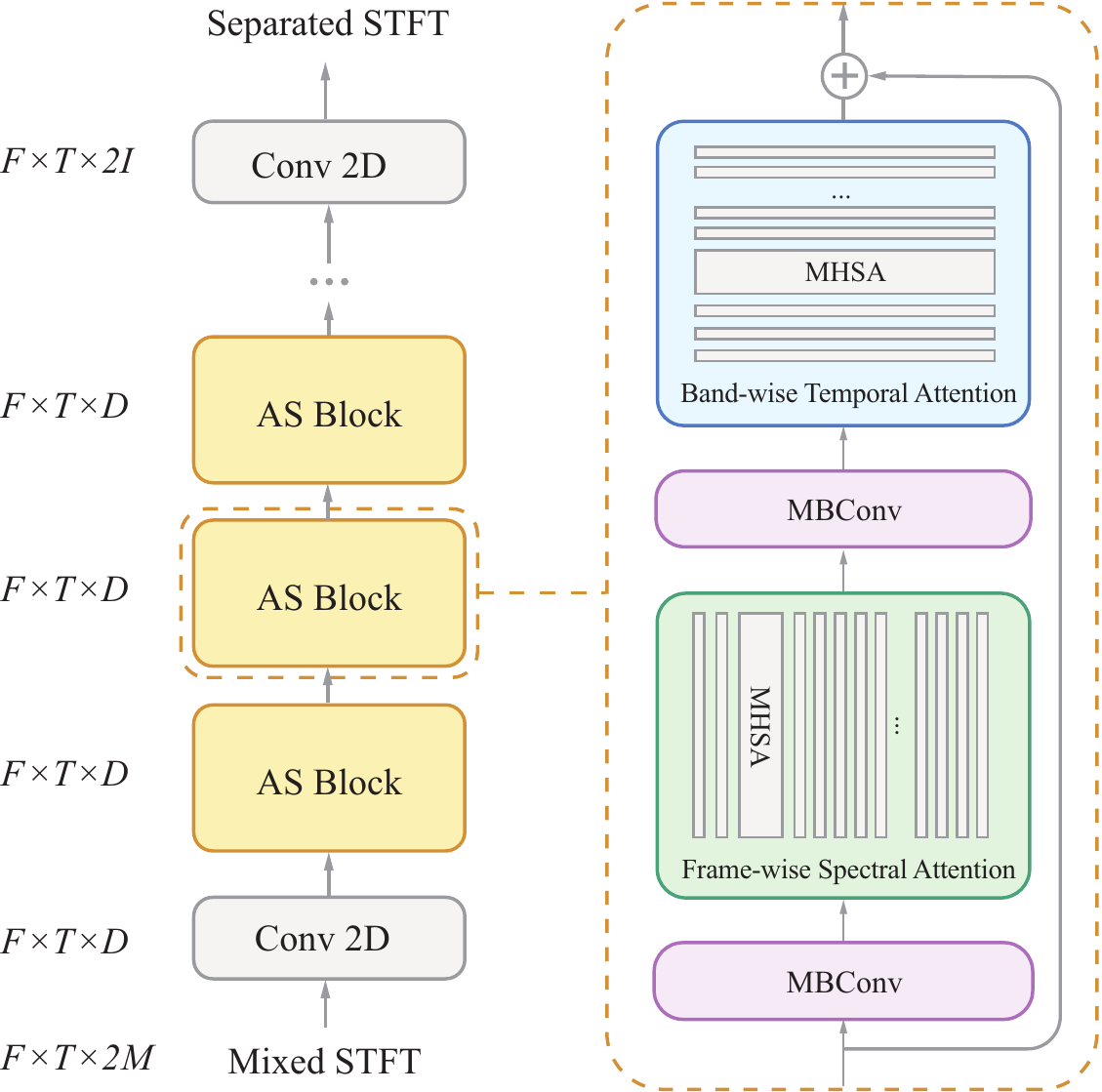} 
		\caption{The architecture of the proposed DasFormer. } 
		\label{Fig.1}
	\end{figure}
	
	\section{The proposed approach}
	\subsection{Formulation}
	
	Assuming that I speakers are recorded by M microphones in a room, the received signal of m-th microphone in frequency domain $Y_m(t, f)$ can be represented as
	\begin{align}
		Y_m(t, f) &= \sum_{i=0}^{I-1}X_{i,m}(t, f)\\
		&= \sum_{i=0}^{I-1}\sum_{l=0}^{L-1}S_i(t-l,f)H_{i,m}(l,f),
	\end{align}
	where $X_{i,m}(t, f)$ denotes the contribution of i-th speech component $S_i(t,f)$ to the m-th microphone, and $H_{i,m}(l,f)$ models the convolution process due to room reflections.
	
	\subsection{DasFormer}
	
	Within a single frame, the spectral regions occupied by one speaker have inter-dependencies between these sub-bands, such as harmonic trains, despite their overlap \cite{parsons1976separation}. We employ the multi-head self-attention (MHSA) to model the sequence globally to seek such dependencies and aggregate the components from the same speaker. This is similar to \cite{yang2022tfpsnet} which also used sequence modeling on TF domain. 
	
	Within a sub-band, frames dominated by the same speaker share a consistent convolution process (as in Eq.(2)), which would be stronger when considering the phase difference introduced by the direct path when multiple channels are available. So MHSA is also applied to temporal sequences. As illustrated by Fig.\ref{Fig.1}, we propose the Alternating Spectrogram Block (AS Block) including two MBConvs, a frame-wise temporal attention module and a band-wise spectral attention module as the basic processing unit. 
	Multiple AS Blocks are then repetitively stacked together as the deep alternating spectrogram transformer (DasFormer). Such deep repetition and the alternating processing pipeline proves to be crucial for obtaining better separation results. Specific design of each component is as followed.
	
	
	%
	
	
	
	
	\subsubsection{Frame-wise Spectral Attention (FSA)}
	Denoting the output of the encoder as TF embeddings, $\mathbf{e}(t,f)$. The sequence $\mathbf{e}(t,\cdot)$ consisting of all sub-bands in frame $t$ is fed into a MHSA module, which expressed as
	\begin{align}
		\mathbf{e}(t,\cdot) \leftarrow \mathbf{e}(t,\cdot) + Dropout(MHSA(LN(\mathbf{e}(t,\cdot))))
	\end{align}
	where $LN$ denotes the Layer Normalization. The above is repeated for all frames and the same MHSA module is shared.
	\subsubsection{Band-wise Temporal Attention (BTA)}
	Similar to the FSA, the sequence consisting of all frames at the f-th sub-band passes through an MHSA, and is denoted as
	\begin{align}
		\mathbf{e}(\cdot,f) \leftarrow \mathbf{e}(\cdot,f) + Dropout(MHSA(LN(\mathbf{e}(\cdot,f))))
	\end{align}
	All sub-bands repeated and share the same MHSA module.
	\subsubsection{MBConv}
	Before each Attention module, we add a 3$\times$3 MBConv block with Squeeze-Excitation (SE) module \cite{howard2017mobilenets}. This is similar to the convolution augmented (FFN) in Conformer \cite{gulati2020conformer} and NBC \cite{quan22b_interspeech}, except that 2D convolution is used here, as follows: 
	\begin{align}
		\mathbf{e}(\cdot,\cdot) \leftarrow \mathbf{e}(\cdot,\cdot) + Pw_2(SE(Dw(Pw_1(BN(\mathbf{e}(\cdot,\cdot))))))
	\end{align}
	where $Pw_1$ and $Pw_2$ are both Point-wise Conv2D, which implement the expansion and shrinkage projections with factor set to 4, respectively. The $BN$ denotes Batch Normalization and $DW$ denotes $3\times3$ Depth-wise Conv2D \cite{howard2017mobilenets}.

	\section{Experiment}
	
	
	\subsection{Setup}
	
	We evaluate DasFormer on both multi/single-channel tasks. 
	
	\noindent{\bf Multi-channel dataset.} The commonly used dataset - spatialized WSJ0-2MIX is selected \cite{wang2018multi}. All utterances are segmented into 4-second lengths and convolved with randomly generated room impulse responses (RIR). The 28,000 RIRs are generated using same parameters with \cite{wang2018multi}, including room size, reverberation time (T60), speaker location, and array geometry. These clips are then mixed in a fully overlapped manner according to \cite{hershey2016deep}. To align with \cite{ochiai2020beam, chen22e_interspeech}, the first microphone is used as reference and only the first 4 of the 8 microphones are used as inputs. The sampling rate is 8 kHz.
	To compare with \cite{quan22b_interspeech}, we added another settings adopted in \cite{quan22b_interspeech}. The main differences include, 8 microphone arrays with fixed geometry, larger T60, 16 kHz sampling rate.
	In two above settings, we focus on pure separation by using reverberated speech as the training target.
	
	\noindent{\bf Single-channel dataset.}
	This is actually a special case when $M=1$, e.g. using only one microphone data. For comparison with reported results, we choose a widely used and more challenging single-channel dataset - WHAMR! \cite{maciejewski2020whamr}, whose goal is to predict each speaker's clean signal from reverberant and noisy input and hence is closer to the real-world scenario.

	\noindent{\bf Model implementations.} The MHSA module in FSA and BTA modules both applies a dimension $D=64$ and number of heads $H=4$. The parameters of the two MBConv modules are: the kernel size of DW Conv is $3\times3$, the number of channels $D=64$, the expansion factor is $4$, and the shrinkage factor in the SE module is $0.25$. The AS Block repeats $L=12$ layers. The initial encoder is a $3\times3$ Conv2D with $2M$ input channels and $D$ output channels. The final decoder is also a $3\times3$ Conv2D with $D$ input channels and $2I$ output channels. The model architecture is consistent by default on both tasks, except for the number of input channels. A frame length of 32 ms with 16 ms frame shift is used in STFT.
	
	\noindent{\bf Training configurations.} An Adam optimizer is used with an initial learning rate of $0.001$. The signal metric SI-SDR is used as the loss function. The learning rate halved when no lower SI-SDR is found for 7 consecutive epochs. When no lower metric is found for 15 consecutive epochs, training stops. Gradient clipping with a maximum norm of 5 is used to avoid gradient explosion.
	\begin{table}[t]
		\small
		\centering
		\begin{tabular}{lccccc}
			\hline \hline Model& Params.  & PESQ & SDRi  \\
			& $(\mathrm{M})$ & & $(\mathrm{dB})$ \\
			\hline \hline RIR settings as \cite{ochiai2020beam, chen22e_interspeech}\\
			\hline Mixture & $-$ & $1.80$ & $0.0$ &  \\
			FaSNet-TAC \cite{luo2020end} & $2.8$ & $2.90$ & $11.7$ & \\
			NBC \cite{quan22b_interspeech} & 2.0 &2.95 &13.3 \\
			Beam-TasNet \cite{ochiai2020beam} & $-$ & $-$ & $16.8$ & \\
			BeamGuided-TasNet \cite{chen22e_interspeech} & $5.4$ & $-$ & $21.5$ & \\
			DasFormer (ours) & $2.2$ & $\mathbf{4.33}$ & $\mathbf{25.9}$ &  \\
			\hline  \hline
			RIR settings as \cite{quan22b_interspeech}\\
			\hline Mixture & $-$ & $1.80$ & $0.0$ &  \\
			NBC \cite{quan22b_interspeech} & $2.0$ & $3.53$ & $15 . 3$ &  \\
			DasFormer (ours) & $2.2$ & $\mathbf{4.11}$ & $\mathbf{20.5}$ &  \\
			\hline 
		\end{tabular}
		\caption{Experiment results on spatialized WSJ0-2Mix.}
		\label{Tab.1}
	\end{table}
	\begin{table}[t]
		\small
		\centering
		\begin{tabular}{lccc}
			\hline
			\hline Model & Params. & SI-SDRi & SDRi \\
			& $(\mathrm{M})$ & $(\mathrm{dB})$ & $(\mathrm{dB})$ \\
			\hline TasNet-BLSTM\cite{maciejewski2020whamr} & $23.6 $ & $9.2$ & $-$ \\
			Conv-TasNet \cite{maciejewski2020whamr} & $8.8 $ & $8.3$ & $-$ \\
			DPRNN \cite{luo2020dual}& $2.6 $ & $10.3$ & $-$ \\
			DPTNET \cite{chen2020dual}& $2.7 $ & $12.1$ & $11.1$ \\
			Sepformer \cite{subakan2021attention}& $26 $ & $11.4$ & $-$ \\
			WaveSplit \cite{zeghidourWavesplitEndtoEndSpeech2021}& $ 29 $ & $12.0$ & $11.1$ \\
			WaveSplit (DM) \cite{zeghidourWavesplitEndtoEndSpeech2021} & $29 $ & $13.2$ & $12.2$ \\
			Sudo rm -rf (U=16) \cite{tzinis2022compute} & $6.3 $ & $12.1$ & $-$ \\
			Sudo rm -rf (U=36) \cite{tzinis2022compute} & $26.6 $ & $13.5$ & $-$ \\
			\hline 
			DasFormer & $2.2$ & $16.0$ & $14 . 6$ \\
			DasFormer Plus & $6.4$ & $\mathbf{17 . 3}$ & $\mathbf{15 . 7}$ \\
			\hline
		\end{tabular}
		\caption{Experiment results on WHAMR!.}
		\label{Tab.2}
	\end{table}
	
	\begin{figure}[t]
		\centering
		\includegraphics[width=8cm]{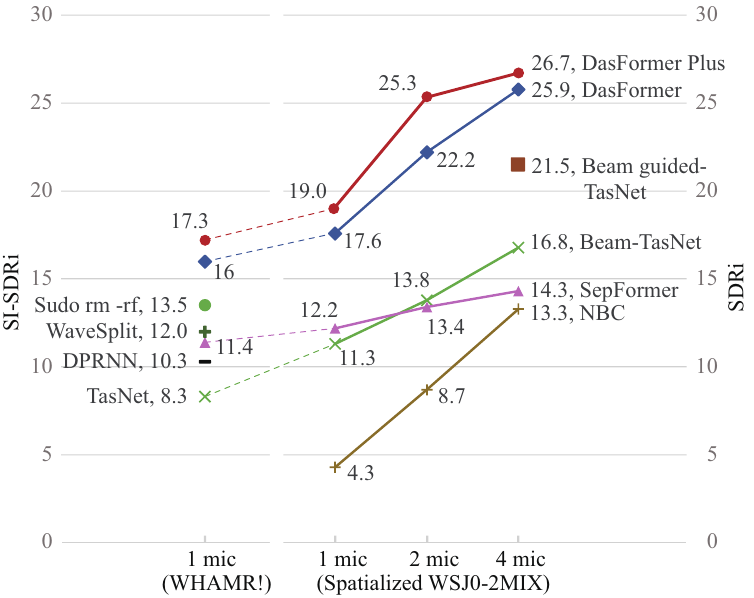}
		\caption{Results on different microphone numbers. All results are on spatialized WSJ0-2MIX with SI-SDRi as metric except for the first column which is on WHAMR! with SDRi metric.} 
		\label{Fig.2}
	\end{figure}
	
	\begin{table}[t]
		\small
		\centering
		\begin{tabular}{cccccc}
			\hline \hline Layers & Dims & Conv & Params. & SDRi  \\
			$(\mathrm{L})$ & $(\mathrm{D})$ & &$(\mathrm{M}$) & $(\mathrm{dB})$ \\
			\hline 12 & 64 & - & $2.2$ & $\mathbf{25.9}$ &  \\
			8 & 80 & - & $2.2$ & $24.1$ &  \\
			4 & 128 & - & $2.2$ & $23.4$ & \\
			\hline 12 & 64 & w/o SE & $1.4$ & $24.3$ &  \\
			12 & 64 & $1\times1$, w/o SE & $1.3$ & $22.2$ &  \\
			\hline 
		\end{tabular}
		\caption{Results on ablation study. '-' means that the MBConv module is not changed. 'w/o SE' means remove the SE module. '$1\times1$' means that the $3\times3$ Conv2D in the MBConv module is replaced with a point-wise Conv2D.}
		\label{Tab.3}
	\end{table}
	
	\subsection{Performance comparison}
	
	{\bf Multi-channel speech separation} We use SDR improvement (SDRi) and narrow-band PESQ to evaluate DasFormer and each baseline system. Our proposed DasFormer achieves an SDR improvement of 25.9 dB with the RIR setting in \cite{wang2018multi}, which is a 4.4 dB improvement compared to BeamGuided-TasNet, an approach with an iterative refinement framework combined with spatial filters and single-channel separation in the time domain. The DasFormer also obtained an SDR improvement of 20.5 dB under the RIR setting in \cite{quan22b_interspeech}, a 5.2 dB improvement over the NBC approach with narrow-band complex mapping. 
	We also notice that the NBC model working in narrowband mode degrades significantly in the task with random array geometry, while DasFormer achieves a higher SDR improvement under this challenging scenario instead. Considering the diversity and phase inconsistency of arrays in practice, this random array geometry setting is a more robust and adaptive way to training models.
	
	
	\noindent{\bf Single-channel separation with reverberation} DasFormer and its plus version (D=96, L=16) achieved SI-SDR improvements of 16.0 dB and 17.3 dB, respectively, which is a 3.8 dB performance improvement compared to the SOTA model on WHAMR! dataset \cite{tzinis2022compute}.  
	
	
	
	\subsection{Model scalability on microphone number}
	We compare DasFormer and current mainstream models on various microphone numbers. The WHAMR! in Fig. \ref{Fig.2} can be viewed as a more difficult single microphone setting with noise added. For aligning, we extend Sepformer by increasing the input channels of encoder. However, from the limited performance gain after increasing channels, it is obvious that the spatial cues are less effectively utilized by SepFormer. And Beam-TasNet is far inferior to other SOTA algorithms when degraded to single-channel (when MVDR is not available). The NBC method is severely degraded, which can be explained by the fact that its narrow-band mode relies heavily on spatial information, which is susceptible to channel reduction. The proposed DasFormer not only achieves the highest performance on both tasks, but also increases the performance significantly as the microphone number increases.
	
	\subsection{Ablation Study}
	To investigate the contribution of deeper layers and the MBConv module, we conducted two sets of ablation experiments (all trained and tested on the spatialized WSJ0-2Mix dataset \cite{wang2018multi}). The first one gradually makes the DasFormer shallower and keeps the model size constant by increasing the embedding dimension. It shows that deeper networks are easier to achieve higher performance for DasFomer. The second one is to remove the SE module and the $ 3\times 3$ Conv from MBConv consecutively. The $ 3\times 3$ Conv is replaced by a point-wise Conv, becoming similar to the FFN module in Transformer. It can be found that both of them bring significant degradation. And the $ 3\times 3$ Conv2D is more crucial for better performance.
	
			
	
	
	\subsection{Visualization}
	\begin{figure}[t]
		\centering
		\includegraphics[width=7.5cm]{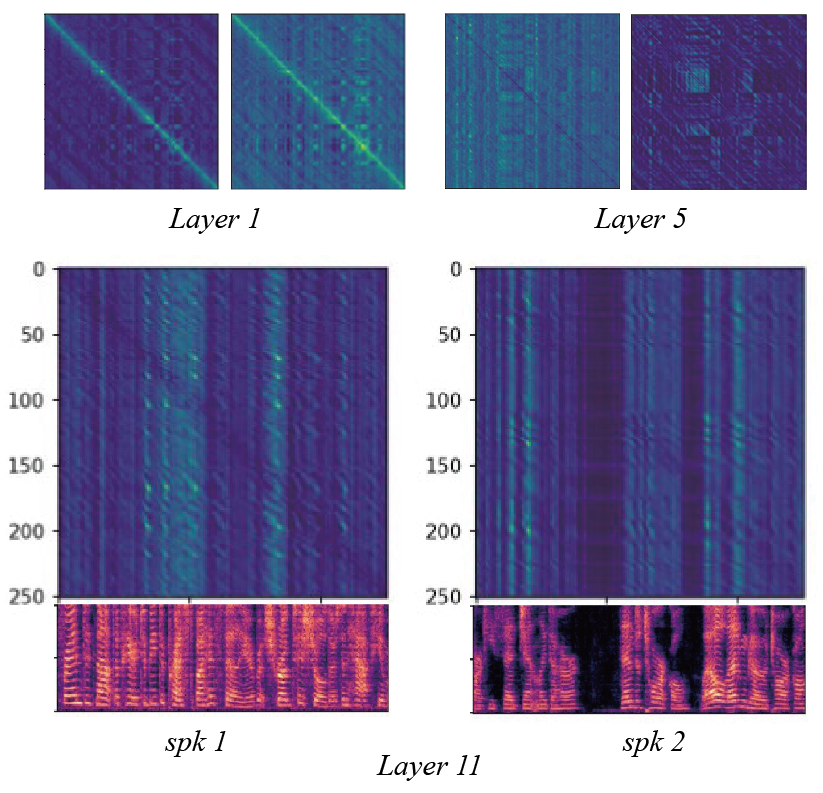}
		\caption{Results of attention scores from different layers in AS Block. They are two heads from the shallow (layer 1), middle (layer 5) and deep (layer 11) layer respectively.} 
		\label{Fig.3}
	\end{figure}
	
	To further understand the behavior of the model, we visualize the attention map. We observe that different layers present different temporal patterns. The shallow layer pays more attention on local scope formed by neighboring frames, as evidenced by higher scores around the diagonal. The middle layer reflects more block-like structure, which may attribute to the similarity of frames within the same phonetic unit. And the heads from deeper layers show distributions with clear correspondence to speaker activity. So as the network becomes deeper, the DasFormer tends to encourage different layers to aggregate different levels of information.
	
	


	\section{Conclusion}
	This work attempt to employ a common architecture for both multi/single-channel speech separation. By alternately performing band-wise and frame-wise MHSA on TF-bin embedding and combining spatial information when multi-channel data are available, the proposed DasFormer can maintain robust separation results against less microphone number and even gets SOTA results on single-channel speech separation in the challenging reverberant environments. The proposed architecture is potential for more challenging scenarios, like adapting a model trained on one microphone array to another (different on both microphone number and array geometry).
	\clearpage

	\clearpage

	\bibliographystyle{IEEEbib}
	\bibliography{refs}
	
\end{document}